\documentclass{article}
\pdfoutput=1
\usepackage{graphicx}  
\usepackage{amsmath}   
\usepackage[compress]{cite}
\usepackage{amssymb}   
\usepackage{bm} 
\usepackage{dcolumn}
\usepackage{color}
\usepackage[caption=false]{subfig}
\usepackage{mathrsfs}
\usepackage{amsfonts}
\usepackage{varioref}
\usepackage{float}
\restylefloat{table}
\usepackage[caption=false]{subfig}
\RequirePackage[colorlinks,citecolor=blue,urlcolor=magenta,linkcolor=blue]{hyperref}
\input epsf
\usepackage[labelsep=none,labelformat=empty]{caption}
\def\gr{general relativity}

\labelformat{section}{Section #1} 
\labelformat{subsection}{Section #1} 
\labelformat{subsubsection}{Section #1}
\labelformat{subsubsubsection}{Section #1}
\labelformat{equation}{Eq.~(#1)} 
\labelformat{figure}{Fig.~#1} 
\labelformat{subfigure}{Fig.~\thefigure#1} 
\labelformat{table}{Table.~#1} 
\labelformat{appendix}{Appendix #1}

\title{Quasar continuum spectrum disfavors black holes with a magnetic monopole charge}
\author{Indrani Banerjee\footnote{banerjeein@nitrkl.ac.in} ~$^{1}$, Vijay Shersingh Chawan\footnote{420ph2055@nitrkl.ac.in}~$^{1}$, Bhaswati Mandal\footnote{tpbm3@iacs.res.in}~$^{2}$, Siddharth Kumar Sahoo \footnote{521ph1007@nitrkl.ac.in}~$^{1}$ and \\ Soumitra SenGupta\footnote{tpssg@iacs.res.in}~$^{2}$\\
{\small{$^{1}$Department of Physics and Astronomy, National Institute of Technology, Rourkela-769008, India}}\\
{\small{$^{2}$School of Physical Sciences, Indian Association for the Cultivation of Science, Kolkata-700032, India
}}}

\begin{document}

\maketitle

\begin{abstract}
Black holes carrying a magnetic monopole charge are a subject of interest for a long time. In this work we explore the possibility of an observational evidence of such black holes carrying a magnetic monopole, namely the Bardeen rotating black holes. We derive the theoretical spectrum from the accretion disk surrounding a Bardeen black hole using the thin-disk approximation. We compare the theoretically derived spectrum with the optical data of eighty Palomar Green quasars to constrain the monopole charge parameter $g$ and the spin parameter $a$ of the quasars. From our analysis we note that the Kerr-scenario in \gr\ is observationally more favored than black holes with a monopole charge. We arrive at such a conclusion using error estimators like $\chi^2$, the Nash-Sutcliffe efficiency, the index of agreement and their modified forms.
In particular, black holes with $g \geq 0.03$ are outside $99\%$ confidence interval. The implications are discussed.

\end{abstract}

\section{INTRODUCTION}\label{S1_Bardeen}
General relativity (GR) is the most successful theory of gravity till date in explaining the nature of spacetime around us. From table top experiments to prediction of planetary motion general relativity has been verified to astounding precision \cite{Will:2005yc,Will:1993ns,Will:2005va}. The two recent ground breaking discoveries namely, the gravitational waves \cite{Abbott:2017vtc,TheLIGOScientific:2016pea,Abbott:2016nmj,TheLIGOScientific:2016src,Abbott:2016blz} and the black hole image \cite{Fish:2016jil,Akiyama:2019cqa,Akiyama:2019brx,Akiyama:2019sww,Akiyama:2019bqs,Akiyama:2019fyp,Akiyama:2019eap} have further validated that general relativity holds good in the strong field regime as well. Although GR continues to be the most successful theory of gravity till date it cannot adequately address the dark sector \cite{Milgrom:1983pn,Milgrom:2003ui,Bekenstein:1984tv,Perlmutter:1998np,Riess:1998cb}. Moreover, the theory fails to explain the black hole and cosmological singularities. This has led to a proliferation of alternative gravity theories to compensate the deficits of GR. Such alternative theories include higher curvature gravity \cite{Nojiri:2003ft,Nojiri:2006gh,Capozziello:2006dj,Lanczos:1932zz,Lanczos:1938sf,Lovelock:1971yv,Padmanabhan:2013xyr}, extra-dimensional models \cite{Shiromizu:1999wj,Dadhich:2000am,Harko:2004ui,Carames:2012gr} and the scalar-tensor/scalar-vector-tensor theories of gravity \cite{Horndeski:1974wa,Sotiriou:2013qea,Babichev:2016rlq,Charmousis:2015txa} which reproduce GR in the low energy limit. 

Black holes in general relativity have a curvature singularity at $r=0$ where the theory itself breaks down. The rotating black hole solution in GR i.e. the Kerr scenario turns out to be astrophysically most relevant. The goal of the present work is to consider modifications to Einstein gravity in light of non-linear electrodynamics. Such non-linear modifications to the electromagnetic action are interesting as they give rise to regular black holes which evades the singularity at $r=0$. In this regard we study the Bardeen black hole scenario in light of astrophysical observations which are associated with a magnetic monopole charge. Such black holes are characterized by the monopole charge parameter $g$ and the spin $a$. The abundance of astrophysical data in the electromagnetic domain gives us a scope to test these alternatives to GR. Since gravity is expected to be the strongest near the horizon of black holes, the near horizon regime of black holes seem to be important astrophysical sites to test various modifications of GR. 

In this paper we use observations related to the continuum spectrum of black holes to discern the signatures of the background metric. The continuum spectrum depends both on the nature of the background spacetime as well as the characteristics of the accretion flow. We consider the thin accretion disk model \cite{Novikov_Thorne_1973,Page:1974he} to derive the continuum spectrum in the Bardeen metric. The theoretically derived continuum spectrum is then compared with the optical data of eighty Palomar Green quasars to extract information about the background metric. We use error estimators, in particular, chi-square analysis to arrive at the observationally favored monopole charge. Our analysis reveals that the Kerr scenario is observationally more favored than black holes with a monopole charge. In particular, monopole charges higher than $g\gtrsim 0.13$ are outside $99\%$ confidence interval. Our results are further confirmed by error estimators like 
Nash-Sutcliffe efficiency, index of agreement and their modified forms.

The paper is organised as follows: In \ref{S2_Bardeen}, we discuss about the Bardeen black hole solution in non-linear electrodynamics. 
The theoretical spectrum from the accretion disk is evaluated in this background metric in \ref{S3_Bardeen}. \ref{S4_Bardeen} is dedicated to comparison of the optical data of eighty PG quasars with the theoretical spectrum and the associated error analysis. We summarize our results with some scope for future work in \ref{S5_Bardeen}.\\
Notations and Conventions: In this paper we use mostly positive metric convention and consider G = c = 1.

\section{Brief introduction to Bardeen rotating black hole}\label{S2_Bardeen}
The general rotating, stationary and axially symmetric black hole metric in Boyer-Lindquist coordinates is,
\begin{align}\label{metric_bardeen}
ds^{2} &=-\bigg{(} 1 - \frac{2\bar{m}(r)r}{\tilde{\Sigma}}\bigg{)}dt^{2} - \frac{4\tilde{a}\bar{m}(r)r}{\tilde{\Sigma}}\sin^{2}\theta dt d\phi + \frac{\bar{\Sigma}}{\Delta}dr^{2}\nonumber\\
&+\bar{\Sigma} d\theta^{2} + \bigg{(} r^{2} + \tilde{a}^{2} + \frac{2\bar{m}(r)r\tilde{a}^{2}}{\tilde{\Sigma}}\sin^{2}\theta\bigg{)}\sin^{2}\theta d\phi^{2}
\end{align}
where, 
\begin{align}\label{metricparams_bardeen}
\bar{\Sigma} = r^{2} + \tilde{a}^{2}\cos^{2}\theta ~ {,} ~ \Delta = r^{2} + \tilde{a}^{2} - 2\bar{m}(r)r
\end{align}
Here, $\bar{m}(r)$ is the mass function. Also $lim_{r\rightarrow\infty}\bar{m}(r) = {\mathcal{M}}$ and $\mathbf{\tilde{a}}$ is the spin parameter which can be written as, $\mathbf{\tilde{a}=\frac{\bar{J}}{\bar{\mathcal{M}}}}$ by definition. $\bar{J}$ represents the angular momentum of the rotating black hole and ${\mathcal{M}}$ represents the ADM mass of the same. Now the metric in \ref{metric_bardeen} reproduces Kerr spacetime when $\bar{m}(r) = {\mathcal{M}}$ and Kerr-Newman spacetime with $\bar{m}(r) = {\mathcal{M}} - \frac{Q^{2}}{2r}$.
\par
Bardeen\cite{PhysRev.174.1559} first proposed regular black hole with horizons and no curvature singularity which is a modification of the renowned Reissner-Nordstrom(RN) black hole. Our interest of the project which is famous rotating Bardeen black hole\cite{BAMBI2013329} belongs to this prototype of non-Kerr black hole family. The mass function of the Bardeen rotating black hole reads as,
\begin{align}\label{massfn}
\bar{m}(r) = {\mathcal{M}}\bigg{[} \frac{r^2}{r^2 + \tilde{g}^{2}}\bigg{]}^{3/2}
\end{align}
Here ${\mathcal{M}}$ is the mass of the black hole and $\tilde{g}$ is the magnetic monopole charge of a self-gravitating magnetic field described by a nonlinear electrodynamics\cite{AYONBEATO2000149}. $\bar{m}(r)$ can be interpreted as the mass inside the sphere of radius $r$. This spacetime is regular everywhere and it satisfies the weak energy condition. This Bardeen black hole depicts a regular space-time with curvature invariants,
\begin{align}
\mathcal{R} &= \frac{6{\mathcal{M}}\tilde{g}^2(4\tilde{g}^{2} - r^{2})}{(r^{2} + \tilde{g}^{2})^{7/2}} \\
\mathcal{R}^{\sigma\rho}\mathcal{R}_{\sigma\rho} &= \frac{18{\mathcal{M}}^{2}\tilde{g}^4(8\tilde{g}^4 - 4\tilde{g}^2 r^2 + 13r^4)}{(r^2 + \tilde{g}^2)^7} \\
\mathcal{R}^{\sigma\rho\delta\alpha}\mathcal{R}_{\sigma\rho\delta\alpha} &= \frac{12{\mathcal{M}}^2(8\tilde{g}^8 - 4\tilde{g}^6 r^2 + 47\tilde{g}^4 r^4 - 12\tilde{g}^2 r^6 + 4r^8)}{(r^2 + \tilde{g}^2)^7}
\end{align}
As stated above Bardeen black hole is similar to RN black hole except the usual singularity of RN solution at $r =0$ which is smoothed out here and corresponds to the origin of spherical coordinates. Since $\tilde{g}$ is associated with nonlinear electrodynamics, the dynamics of the theory is governed by the action,
\begin{align}\label{action_bardeen}
\mathcal{S} = \int ~ d^4x \sqrt{-\eta}\bigg{[} \frac{\mathcal{R}}{16\pi} - \frac{\mathcal{W}(f)}{4\pi}\bigg{]}
\end{align}
with $\mathcal{R}$ as scalar curvature and $\mathcal{W}(f)$ is a function of $f = \frac{1}{4}\mathcal{F}_{\mu\nu}\mathcal{F}^{\mu\nu}$ where $\mathcal{F}_{\mu\nu} = 2\nabla_{[\mu}A_{\nu]}$ as the electromagnetic field strength. The action \ref{action_bardeen} results into the Einstein nonlinear electrodynamics field equations as,
\begin{align}
&\mathcal{G}^{\nu}_{\mu} = 2\bigg{[}\mathcal{W}_{f}(\mathcal{F}_{\mu\lambda}\mathcal{F}^{\nu\lambda}) - \delta^{\nu}_{\mu}\mathcal{W}\bigg{]}\\
&\nabla_{\mu} (\mathcal{W}_{f}\mathcal{F}^{\beta\mu}) =0
\end{align}
with $\mathcal{W}_{f} = \frac{\partial\mathcal{W}}{\partial f}$. The particular form of the function $\mathcal{W}$ representing the Bardeen black hole is as follows,
\begin{align}
\mathcal{W}(f) = \frac{3\mathcal{M}}{|\tilde{g}|\tilde{g}^2}\bigg{[} \frac{\sqrt{2\tilde{g}^2 f}}{1 + \sqrt{2\tilde{g}^2 f}}\bigg{]}^{5/2}
\end{align}
It is important to note that Kerr black hole can be retrieved in the absence of nonlinear electrodynamics i.e $\tilde{g} = f=0$. We aim to constrain the both the parameters $a$ and $\tilde{g}$ in the light of astrophysical observations.
\par
The event horizon of this said spacetime is evaluated from $g^{rr} = 0 \Rightarrow \Delta = 0$ which leads to solving the equation,
\begin{align}\label{horizon_bardeen}
r^2 + \tilde{a}^2 - 2\mathcal{M}r\bigg{(} \frac{r^2}{r^2 + \tilde{g}^2}\bigg{)}^{3/2}=0
\end{align}
We aim to solve this for real and positive event horizons of the black hole and constrain the magnetic monopole charge parameter $g = \tilde{g}^2/\mathcal{M}^2$. Now we will give a prescription to demonstrate the dependence of luminosity from the accretion disk on the background spacetime.

\section{ACCRETION DISK AROUND ROTATING BARDEEN BLACK HOLE}\label{S3_Bardeen}
In this section we derive the continuum spectrum emitted from the accretion disk around a rotating black hole (which gives rise to a stationary, axi-symmetric background spacetime) in the thin-disk approximation \cite{Novikov_Thorne_1973,Page:1974he}. The continuum spectrum emitted by the accretion flow around a black hole depends on the background metric and also on the characteristics of the accretion flow. Therefore, the continuum spectrum is an important observational tool to discern the signatures of the background spacetime. In the thin-disk approximation, the accretion flow is assumed to be localized in the equatorial plane where $\theta = \pi/2$ such that the disk height $h(r)\ll r$ where $r$ is the radial distance from the central black hole. 
The azimuthal velocity of the accreting particles $u_{\phi}$ is much greater than the radial velocity $u_{r}$ and vertical velocity, $u_{z}$ such that $u_{\phi} \gg u_{r} \gg u_{z}$ and this ensures nearly circular geodesics of the particles. The viscous stress transmits minimal radial velocity to the accreting fluid such that the accreting matter slowly inspirals and falls into the black hole. The thin disk harbors no outflows as the vertical velocity is negligible. The energy-momentum tensor of the accreting fluid can be written as,
\begin{align}\label{EM_bardeen}
\mathcal{T}^{\mu}_{\nu} = \rho_{0}(1 + \tilde{\Pi})u^{\mu}u_{\nu} + t^{\mu}_{\nu} + u^{\mu}q_{\nu} + q^{\mu}u_{\nu}
\end{align}

In the above expression, $u_{\nu}$ is the four velocity and $\rho_{0}$ is the proper density of the accreting particles. $t^{\mu\nu}$ is the stress-tensor and $q^{\mu}$ is the energy flux relative to the local inertial frame such that $t_{\mu \nu}u^{\mu}=0=q_{\mu}u^{\mu}$. Also $\tilde{\Pi}$ denotes the specific internal energy of the accreting fluid and the associated term expresses the contribution to the energy density due to dissipation. In the thin-disk approximation $\tilde{\Pi} \ll 1$ (i.e no heat is retained by the accreting fluid) such that only $z-$ component of the energy flux vector $q^{z}$ has effective non-zero contribution to the energy-momentum tensor.
The specific internal energy of the accreting fluid is assumed to be negligible in comparison with its rest energy (i.e. $\tilde{\Pi}\ll 1$) which ensures that the special relativistic corrections due to local thermodynamic, hydrodynamic and radiative properties of the fluid can be neglected in comparison to its rest energy. This in turn ensures that the accretion flow remains geodetic. However, the general relativistic corrections associated with the black hole continues to be significant \cite{Novikov_Thorne_1973,Page:1974he}. 
The photons thus emitted as a result of viscous dissipation interact effectively with the accreting fluid before reaching the observer as a result of which each annulus of the accreting disk emits black body radiation. Therefore, the continuum spectrum from the thin accretion disk in the Novikov-Thorne approximation turns out to be a multi-color black body spectrum. 

\par
After discussing the model approximation, next we will compute the flux and then the luminosity from the accretion disk. The black hole accretes at a steady rate, $\dot{M}$ and hence the accreting fluid follows mass conservation, energy conservation and angular momentum conservation laws.
\begin{itemize}
\item{\textbf{Mass Conservation:}
\begin{align}\label{MassCons_bardeen}
\dot{M} = -2\pi\sqrt{-\eta}u^{r}\tilde{\sigma}
\end{align}
Here $\tilde{\sigma}$ is the average surface density of accreting matter falling into the black hole. The determinant of the metric is $\eta = -g_{rr}(g^{2}_{t\phi} - g_{tt}g_{\phi\phi})$ as we are focussing on the motion along $\theta = \pi/2$ plane i.e. near- equatorial plane.}
\item{\textbf{Energy Conservation:}
\begin{align}\label{ECons_bardeen}
\partial_{r}(\dot{M}\tilde{\mathcal{L}} - 2\pi\sqrt{-\eta}w^{r}_{\phi})~ = ~ 4\pi\sqrt{-\eta}F\tilde{\mathcal{L}}
\end{align}
}
\item{\textbf{Angular Momentum Conservation:}
\begin{align}\label{AMCons_bardeen}
\partial_{r}(\dot{M}\tilde{\mathcal{E}} - 2\pi\sqrt{-\eta}\tilde{\Omega} w^{r}_{\phi}) ~ = ~ 4\pi\sqrt{-\eta}F\tilde{\mathcal{E}}
\end{align}
}
\end{itemize}

In the above expressions \ref{ECons_bardeen} and \ref{AMCons_bardeen} the unknown parameters are as following, $\tilde{\Omega}$ is the angular velocity, $\tilde{\mathcal{E}}$ is the specific energy and $\tilde{\mathcal{L}}$ is the specific angular momentum of the accreting fluid approximated as test particles. The function $F$ denotes the flux radiated from the accretion disk and given as,
\begin{align}\label{FluxF_bardeen}
F \equiv \langle q^{z}(r,h)\rangle = \langle -q^{z}(r,-h)\rangle
\end{align}
and $w^{r}_{\phi}$ is related to the time and height averaged stress tensor in the local rest frame of the accreting particles and denoted as,
\begin{align}\label{W_bardeen}
w^{\alpha}_{\beta} = \int^{h}_{-h} dz \langle t^{\alpha}_{\beta} \rangle
\end{align}
In a stationary, axi-symmetric background spacetime, $\tilde{\mathcal{E}}$ and $\tilde{\mathcal{L}}$ are conserved quantities which can be written in terms of metric coefficients as,
\begin{align}
\tilde{\mathcal{E}} &= \frac{-g_{tt} - \tilde{\Omega}g_{t\phi}}{\sqrt{-g_{tt} - 2\tilde{\Omega}g_{t\phi} - \tilde{\Omega}^2 g_{\phi\phi}}} \\
\tilde{\mathcal{L}} &= \frac{\tilde{\Omega}g_{\phi\phi} + g_{t\phi}}{\sqrt{-g_{tt} - 2\tilde{\Omega} g_{t\phi} - \tilde{\Omega}^{2} g_{\phi\phi}}}
\end{align}
with the angular velocity, $\tilde{\Omega} = \frac{d\phi}{dt} = \frac{-g_{t\phi,r} \pm \sqrt{(-g_{t\phi,r})^2 - (g_{\phi\phi,r})(g_{tt,r})}}{g_{\phi\phi,r}}$. It is clearly seen that the expressions of $\tilde{\mathcal{E}}$ and $\tilde{\mathcal{L}}$ are only radial functions as we are dealing with motion along the equatorial plane and hence $g_{\theta \theta}$ does not contribute to the conserved quantities.
\par
The conservation laws lead to an analytical expression for the flux $F$ \cite{Page:1974he} and can be expressed as,
\begin{align}
F &= \frac{\dot{M}}{4\pi\sqrt{-\eta}}f \label{FluxForm_bardeen}~~~~\rm {where} \\
f &=~ - \frac{\tilde{\Omega}_{,r}}{(\tilde{\mathcal{E}} - \tilde{\Omega}\tilde{\mathcal{L}})^2} \bigg{[} \tilde{\mathcal{E}}\tilde{\mathcal{L}} - \tilde{\mathcal{E}}_{ms} \tilde{\mathcal{L}}_{ms} - 2\int^{r}_{r_{ms}} \tilde{\mathcal{L}}\tilde{\mathcal{E}}_{,r^{\prime}} dr^{\prime} \bigg{]} \label{f_form_bardeen}
\end{align}
$\tilde{\mathcal{E}}_{ms}$ and $\tilde{\mathcal{L}}_{ms}$ in the expression \ref{f_form_bardeen} refers to the energy and angular momentum of the marginally stable circular orbit $r_{ms}$. In the Novikov-Thorne model the viscous stress $w^{r}_{\phi}$ is assumed to vanish at the marginally stable circular orbit in order to derive \ref{FluxForm_bardeen}.
Hence the azimuthal velocity of the accreting particles vanish after crossing $r_{ms}$ and radial accretion takes over. Now to determine $r_{ms}$, we need to determine the point of inflection of the effective potential in which the accreting fluid moves i.e. by solving $V_{eff} = \partial_{r}V_{eff} = \partial_{r}^{2}V_{eff} = 0$ and the functional form of this potential is expressed as\cite{PhysRevD.100.044045},
\begin{align}\label{potential_bardeen}
V_{eff}(r) ~ = ~ \frac{\tilde{\mathcal{E}}^2 g_{\phi\phi} + 2\tilde{\mathcal{E}} \tilde{\mathcal{L}} g_{t\phi} + \tilde{\mathcal{L}}^2g_{tt}}{g_{t\phi}^2 - g_{tt}g_{\phi\phi}} ~ - ~ 1
\end{align}
It is important to check the vertical stability of the orbit which is discussed in \cite{Ono:2016lql}.
\par
Since the outgoing photons undergo repeated collisions with the accreting particles before being emitted from the system the integrated emission from the accretion disk is a multi-temperature blackbody spectrum.
This ensures that the accretion disk remains `\textit{optically thick and geometrically thin}' and its temperature profile is given by Stefan-Boltzmann law, $T(r)=\left(\mathcal{F}(r)/\sigma\right)^{1/4}$, where $\sigma$ is the Stefan-Boltzmann constant and $\mathcal{F}(r)=F(r)c^6/G^2M^2$ where $F(r)$ can be evaluated from \ref{FluxForm_bardeen} and \ref{f_form_bardeen}.

Accordingly at every radius with temperature profile $T(r)$ the accretion disk radiates Planck spectrum such that the luminosity from the accretion disk at an observed frequency $\nu$ is expressed as,
\begin{align}\label{luminosity_bardeen}
L_{\nu} = 8\pi^2 r_{g}^2 \cos{i} \int^{x_{out}}_{x_{ms}} \sqrt{g_{rr}} P_{\nu}(T)x~ dx
\end{align}
where $x = r/r_g$ is the radial coordinate with gravitational radius, $r_g = GM/c^2$, $i$ is the inclination angle between the line of sight of the observer and the normal to the accretion disk and
\begin{align}\label{planckfn_bardeen}
P_{\nu} (T) = \frac{2h\nu^3}{c^2\big{[} exp(\frac{h\nu} { z_{g}kT}) -1\big{]}}
\end{align}
In \ref{planckfn_bardeen}, $z_g$ is the gravitational redshift factor corresponding to the outgoing photons and is expressed as,
\begin{align}\label{redshift_bardeen}
z_{\rm g}=\mathcal{E}\frac{\sqrt{-g_{tt}-2\tilde{\Omega} g_{t\phi}-\tilde{\Omega}^2 g_{\phi\phi}}}{\mathcal{E}-\Omega \mathcal{L}} 
\end{align}
In \ref{redshift_bardeen}, the specific energy, angular momentum and angular velocity of the outgoing photon is $\mathcal{E}$, $\mathcal{L}$ and $\Omega$. While travelling from emitting material to the observer, photon experiences change in frequency and is given by the redshift factor  \cite{Ayzenberg:2017ufk}. From \ref{luminosity_bardeen} it is evident that the luminosity from the disk $L_{\nu}$ is dependent not only on the mass of the black hole, accretion rate and the inclination angle but also on the background metric through the marginally stable circular orbit radius, energy, angular momentum and the angular velocity of the accretion disk.
\begin{figure}[t!]
\begin{center}
\hspace{-2.1cm}
\includegraphics[scale=0.69]{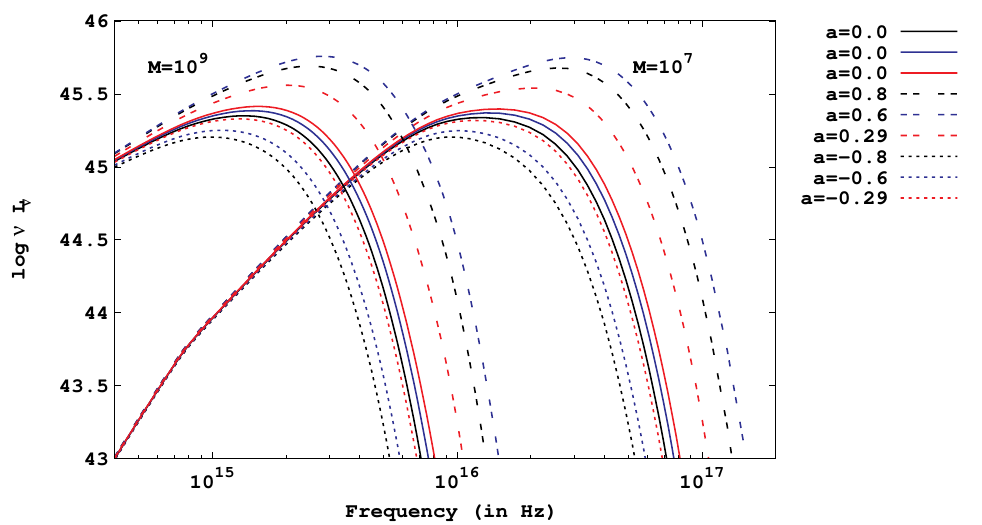}
\end{center}
\caption{Figure 1: In this figure we plot the theoretical luminosity from the accretion disk with frequency assuming black hole masses $M=10^9 M_\odot$ and $M=10^7 M_\odot$. 
The red, blue and black lines are associated with $g=0.5$, $g=0.3$ and $g=0$, where $g$ denotes the magnetic monopole charge. Keeping $g$ fixed, non-rotating black holes are represented by solid lines, while dashed and dotted lines correspond to prograde and retrograde black holes. We consider the inclination angle to be $ i=cos^{-1}0.8$ while the accretion rate taken to be $1 M_\odot~ \rm year^{-1}$. 
}
\label{spectra_Bardeen}
\end{figure}

The variation of the theoretically derived luminosity from the disk with the frequency is illustrated in \ref{spectra_Bardeen} for two sets of black holes masses namely $\mathcal{M}=10^7 M_\odot$ and $\mathcal{M}=10^9 M_\odot$.
The accretion rate is assumed to be $1 M_\odot~ \rm year^{-1}$ and the inclination angle is taken to be $\cos i=0.8$. The temperature dependence, $T \propto \mathcal{M}^{-1/4}$ of a multicolor black body with black hole mass $\mathcal{M}$ ensures that the spectrum peaks at a higher frequency from a lower mass black hole \cite{Frank:2002}. For a black hole with a given mass $\mathcal{M}$ the spectrum from the inner disk generally encodes the information of the background metric. Since the inner disk emits higher frequencies compared to the outer region, the high frequency part of the continuum spectrum is useful to decipher the signatures of the metric parameters $g=\tilde{g}/\mathcal{M}$ and $a=\tilde{a}/\mathcal{M}$. 
The spectrum \ref{spectra_Bardeen} is illustrated for three sets of magnetic monopole charge parameters i.e black lines corresponding to $g=0$, blue lines for $g=0.3$ and red lines for $g=0.5$. These choices of $g$ and $a$ are taken from the consideration of a real, positive event horizon \cite{Kumar:2018ple}. 

For each charge parameter we consider solid lines corresponding to non-rotating black holes, prograde spins are represented by dashed lines while retrograde spins correspond to dotted lines.

\section{NUMERICAL ANALYSIS }\label{S4_Bardeen}

In this section we use thin-disk approximation for the accretion phenomena to evaluate the optical luminosities of a sample of eighty Palomar Green (PG) quasars considered in Davis \& Laor \cite{Davis:2010uq}. According to Davis \& Laor \cite{Davis:2010uq}, the observations corresponding to 4861\AA\ is reported which lies in the optical domain.
The masses of all the quasars are constrained independently by reverberation mapping method \cite{Kaspi:1999pz,Kaspi:2005wx,Boroson:1992cf,Peterson:2004nu} while for some of the quasars masses based on $M-\sigma$ method is also reported in \cite{Davis:2010uq}. In this work masses based on reverberation mapping will be considered. The bolometric luminosities of the said quasars are also determined independently using observed data in different domains such as in the optical \cite{1987ApJS...63..615N}, UV \cite{Baskin:2004wn}, far-UV \cite{Scott:2004sv}, and soft X-ray \cite{Brandt:1999cm}. The dominant contribution to the error in the bolometric luminosity comes from the far-UV regime since the the uncertainty in the UV luminosity surpassed all other sources of error. Hence, although the emission from the accretion disk for quasars peaks in the optical or UV regime of the spectrum, it is difficult to unravel the role of the metric from UV observations. Therefore we use data in the optical domain and compare the optical data of quasars with the theoretical estimates as discussed in the previous section.
\par
The accretion disks associated with the quasar sample considered here are considered to be nearly face-on systems such that the inclination angle varies between $\cos i \in \left(0.5,1\right)$ \cite{Antonucci:1993sg,Davis:2010uq,Wu:2013zqa}.
This is in agreement with \cite{2017Ap&SS.362..231P} where the inclination angle for some of the quasars in the sample
have been estimated using degree of polarisation of the scattered radiation from the accretion disk. In this work we allow the inclination angle of the quasars to vary in the aforesaid range.

\par
The observed optical luminosity and the accretion rates are reported in \cite{Davis:2010uq}. These accretion rates are determined using stellar-atmosphere-like assumptions of the disk structure (termed as TLUSTY model) with spin $a = 0.9$ as their base model. However, the amount of variation in the accretion rate is also estimated using  different disk models or black hole spins. If TLUSTY model with spin $a = 0$ is used the accretion rates of quasars with higher mass get enhanced by $40\%$ while that of lower mass get enhanced by $10\%$ compared to the base model. In contrast, with black body models and $a = 0.9$ the accretion rates turn out to be lowered by $10\% - 20\%$ for all quasars. Similarly, if we consider black body model with $a = 0$, quasars with higher mass tends to increase accretion rate by $40\%$ whereas the quasars with lower mass decreases accretion rate by $20\%$ compared to the base TLUSTY model. Hence, no matter what disk model or black hole spin is chosen, the accretion rate varies between $80\%$ to $140\%$ for all the Palomar Green quasars as discussed in \cite{Davis:2010uq}. \\
We determine the theoretical estimates of the optical luminosity $L_{opt}$ by varying the inclination angle and the accretion rates in the aforementioned range which in turn is compared with the observed optical luminosities to obtain a constrain on the metric parameters.
In order to have a real, positive horizon there arises an upper bound on the magnetic monopole charge parameter $g\sim 0.55$ determined by $\Delta=0$ from \ref{metric_bardeen}. We adopt the following procedure to derive an estimate on the most favored choice of $g$ which is outlined below:\\

\begin{enumerate}
\item First we fix a value of $g$ between $0\leq g \leq 0.55$. This constrains the allowed values of spin such that a real positive event horizon exists. For every $g$ we choose an allowed spin $a$. 

\item Keeping $g$ and $a$ fixed we vary $\dot{M}$ in the range $0.8-1.4$ times the accretion rate provided in \cite{Davis:2010uq}. Now, 
for every $\dot{M}$ in the aforesaid range we vary the inclination angle in the range $\cos i \in \left(0.5,1\right)$. 

\item For every combination of $\dot{M}$, $\cos i$ and $a$ the theoretical optical luminosity is calculated for the fixed $g$ at the wavelength 4861\AA. The values of $\dot{M}$, $\cos i$ and $a$ that best reproduces the observed optical luminosity, is considered to be the most favored magnitude of accretion rate, inclination angle and spin for the chosen quasar at the given $g$. We denote these values of accretion rate, spin and inclination angle by $\dot{M}_{min}$, $a_{min}$ and $cosi_{min}$. 
\item Keeping $g$ fixed, this process is repeated for all the eighty PG quasars which gives us the most favored $\dot{M}$, $\cos i$ and $a$ for the chosen $g$. 
\item We now vary $g$ and repeat the above procedure. 

\end{enumerate}
It is important to note that in the above analysis we assume that all the eighty quasars have the same average magnetic monopole charge $g$. Since the magnetic monopole charge varies in a relatively small range $0\leq g \leq 0.55$ this assumption may be justified.
Now to arrive at the most favorable magnetic monopole charge parameter $g$ several error estimators are examined which will be discussed thereafter. 

\subsection{Discussion on Error Estimators}\label{S4_Bardeen-1}

\begin{itemize}
\item {\textbf {Chi-square} $\boldsymbol {\chi^{2}}~$}:~
If  $\{ \mathcal{O}_{i}\}$ represents a set of observed data with possible errors $\{ \sigma_{i} \}$, and \\ $\left\lbrace \Omega_{i}\left(g,a_{min}, \cos i_{min},\dot{M}_{min}\right)\right\rbrace$ denotes the corresponding theoretical estimates of the optical luminosity for a given $g$, then the $\chi^{2}$ of the distribution is given by,

\begin{align}
\label{eq_chi2_Bardeen}
\chi ^{2}\bigg(g,\left \lbrace a_{min}, \cos i_{min},\dot{M}_{min}\right\rbrace \bigg)= \sum _{j}\frac{\left\{\mathcal{O}_{j}-\Omega_{j} \bigg(g, a_{min}, \cos i_{min},\dot{M}_{min}\bigg) \right\}^{2}}{\sigma _{j}^{2}}.
\end{align}

As discussed earlier, $ a_{min}, \cos i_{min}, \dot{M}_{min}$ denotes the set of spin, inclination angle and accretion rate associated with the $j^{th}$ quasar that minimizes the error between the theoretical and the observed optical luminosity for the chosen $g$. For our sample, the error $\{ \sigma_{i} \}$ corresponding to optical luminosities of individual quasars are not explicitly reported in \cite{Davis:2010uq}. Hence in \ref{eq_chi2_Bardeen}, we consider the errors in the bolometric luminosity as the maximum possible error in the estimation of optical luminosity.

\begin{figure}[t!]
\begin{center}
\hspace{-2.1cm}
\includegraphics[scale=0.67]{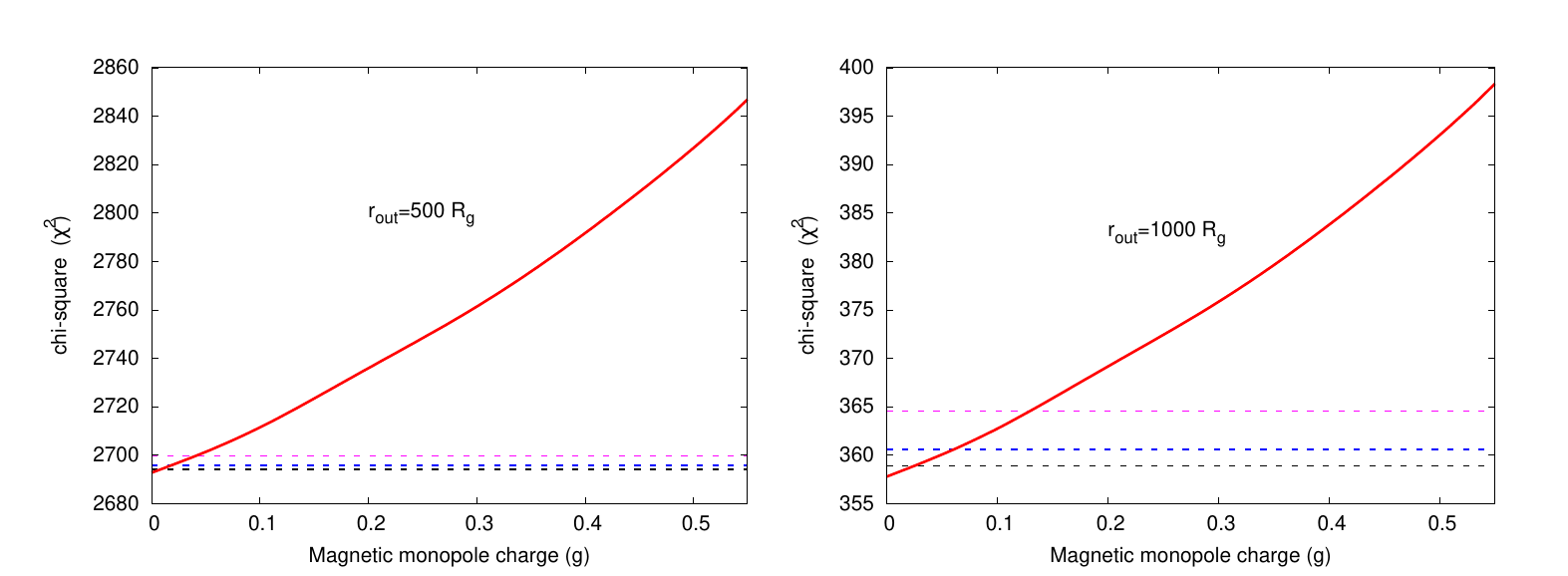}
\end{center}
\caption{Figure 2: The above figure depicts the variation of $\chi^2$ with the magnetic monopole charge $g$ for the sample of eighty PG quasars. In the left panel, the theoretical luminosity and hence the $\chi^2$ is calculated with $r_{out}=500R_g$ while $r_{out}=1000R_g$ is used to compute $\chi^2$ in the right panel. Interestingly, the $\chi^2$ minimizes for the same monopole parameter ($g=0$) irrespective of the choice of $r_{out}$, although the values of $\chi^2$ vary in different range if a different choice of $r_{out}$ is considered. The $\Delta \chi^2$ corresponding to 68\%, 90\% and 99\% confidence intervals (for a single parameter) are also plotted in the figure with black, blue and magenta dashed lines, respectively. }
\label{chi2_Bardeen}
\end{figure}

From \ref{eq_chi2_Bardeen}, we note that the value of $g$ for which $\chi^2$ gets minimized is the most favored value of the monopole 
charge. However, it is important to note that although we have two metric parameters $g$ and $a$ we do not minimize $\chi^2$ for both 
the metric parameters simultaneously since our goal is to investigate the most favored magnetic monopole charge from the quasar data. Therefore, $\chi^2$ is minimized with respect to only one parameter such that $\Delta \chi^2=1, 2.71, 6.63$ corresponding to 
68\%, 90\% and 99\% confidence intervals \cite{1976ApJ...210..642A}. 
In \ref{chi2_Bardeen}, we show the variation of $\chi^2$ with the magnetic monopole charge parameter $g$. Since the theoretical optical luminosity  depends on both the inner radius (which in the present context corresponds to $r_{ms}$) and the outer radius $r_{out}$ we use two sets of outer radii $r_{out}=500R_g$ and $r_{out}=1000R_g$. 
The figure clearly depicts that $\chi^2$ get minimized for $g=0$ irrespective of the choice of outer radius of the disk. The confidence intervals corresponding to 68\%, 90\% and 99\% are shown by the black, blue and magenta dashed lines. We note that when $r_{out}\simeq 500 R_g$, $g\gtrsim 0.03$ is outside 99\% confidence interval while when $r_{out}\simeq 1000R_g$, $g\gtrsim 0.13$ is outside 99\% confidence interval. This shows that very high values of monopole charge are \emph{not favored} from quasar optical data. 
We next discuss several error estimators to validate our findings.

\item \textbf{Nash-Sutcliffe Efficiency and its modified form:}
\begin{figure}
\begin{center}
\hspace{-2.1cm}
\includegraphics[scale=0.67]{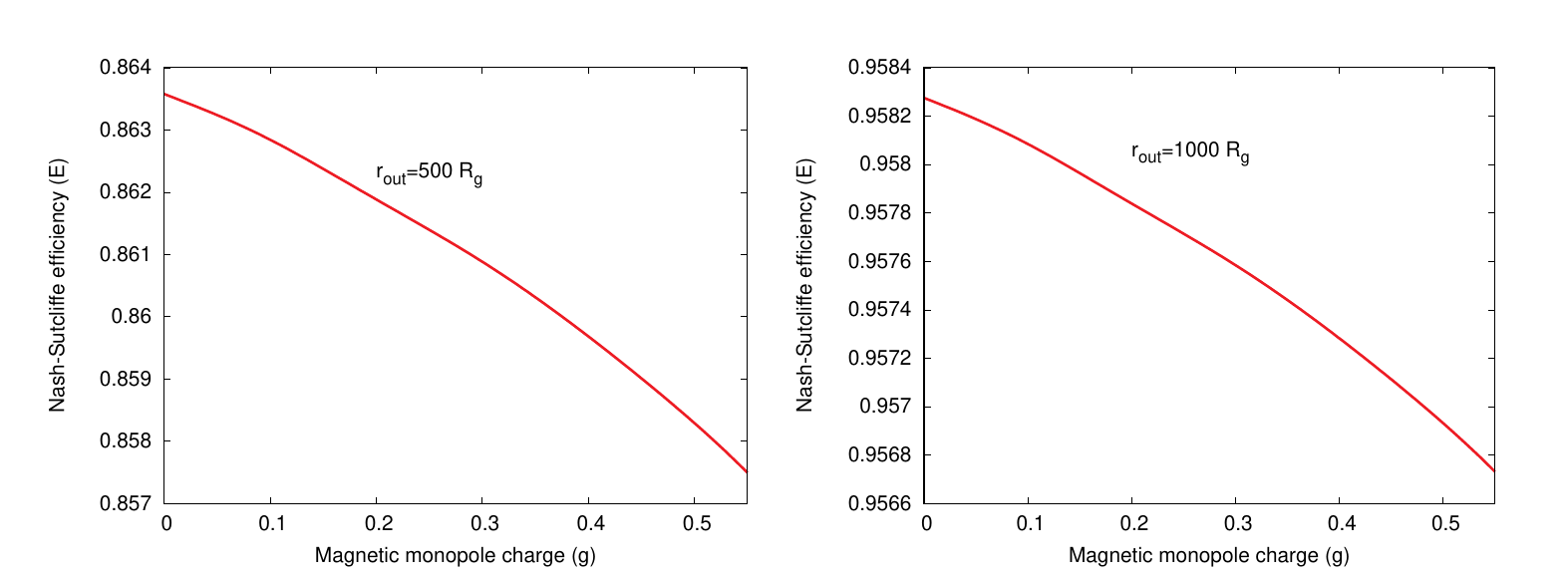}
\end{center}
\caption{Figure 3: The above figure depicts the variation of the Nash-Sutcliffe Efficiency E with the magnetic monopole charge parameter $g$. The Nash-Sutcliffe Efficiency computed with $r_{out}=500r_g$ is shown in the left panel while $E$ computed with $r_{out}=1000r_g$ is shown in the right panel. We note that irrespective of the choice of $r_{out}$, $E$ maximizes for $g=0$.}
\label{NSE_Bardeen}
\end{figure}
The functional form of the Nash-Sutcliffe Efficiency $E$ \cite{NASH1970282,WRCR:WRCR8013,2005AdG.....5...89K} is given by,
\begin{align}
\label{eq_NSE_Bardeen}
E\bigg(g,\left \lbrace a_{min}, \cos i_{min},\dot{M}_{min} \right\rbrace \bigg)=1-\frac{\sum_{j}\left\{\mathcal{O}_{j}-\Omega_{j}\bigg(g,\left \lbrace a_{min}, \cos i_{min},\dot{M}_{min} \right\rbrace \bigg)\right\}^{2}}{\sum _{j}\left\{\mathcal{O}_{j}-\mathcal{O}_{\rm av}\right\}^{2}}
\end{align}
Here, $\mathcal{O}_{\rm av}$ represents the mean value of the observed optical luminosities of the PG quasars. It is important to note that the magnetic monopole charge parameter $g$ for which $E$ gets maximized corresponds to the observationally favored value of $g$. 
From \ref{eq_NSE_Bardeen} it is clear that $E$ can range from $-\infty ~\rm to ~ 1$. A model with negative $E$ indicates that the average of the observed data is a better predictor than the  theoretical model. \ref{NSE_Bardeen} illustrates the variation of Nash-Sutcliffe efficiency $E$ with the charge parameter $g$ for two sets of $r_{out}$. The figure clearly shows that $E$ gets maximized at $g=0$ which is consistent with our finding based on the $\chi^2$ estimate.

\par
Due to the presence of the squared term in the numerator Nash-Sutcliffe Efficiency $E$ is oversensitive to higher values of the luminosity. Therefore, a modified version of the same is proposed which is denoted by $E_1$ \cite{WRCR:WRCR8013}. This is due to the presence of the square of the error in the numerator in \ref{eq_NSE_Bardeen}. Hence, the modified Nash-Sutcliffe Efficiency $E_1$ is expressed as,
\begin{align}\label{eq_E1_Bardeen}
E_{1}\bigg(g,\left \lbrace a_{min}, \cos i_{min},\dot{M}_{min} \right\rbrace \bigg)&=1-\frac{\sum_{j}\big|\mathcal{O}_{j}-\Omega_{j}\bigg(g,\left( a_{min}, \cos i_{min},\dot{M}_{min} \right)\big|}{\sum _{j}\big|\mathcal{O}_{j}-\mathcal{O}_{\rm av}\big|}
\end{align}

\begin{figure}
\begin{center}
\hspace{-2.1cm}
\includegraphics[scale=0.67]{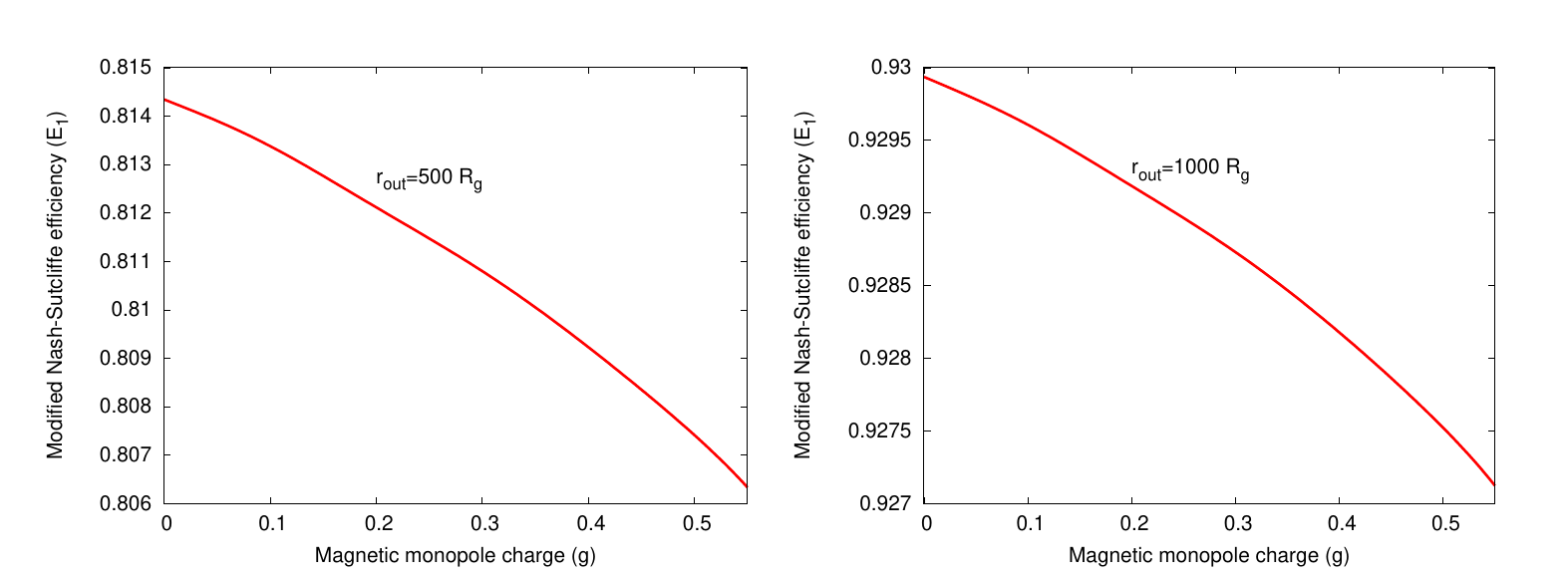}
\end{center}
\caption{Figure 4: The above figure shows the variation of the modified Nash-Sutcliffe Efficiency $E_1$ with the magnetic monopole charge parameter $r_2$. As before, $E_1$ computed with $r_{out}=500r_g$ is shown in the left panel while $E_1$ computed with $r_{out}=1000r_g$ is illustrated in the right panel. As reported in \ref{NSE_Bardeen}, irrespective of the choice of $r_{out}$, $E_1$ maximizes for $g=0$. 
}
\label{E1_Bardeen}
\end{figure}

The figures explicitly elucidate that the most favored $g$ maximizes its modified form as well. \ref{E1_Bardeen} clearly depicts that this maximization appears at $g=0$ irrespective of the choice of outer radius which again validates our earlier findings.

\item \textbf{Index of agreement and its modified form:} 
\begin{figure}
\begin{center}
\hspace{-2.1cm}
\includegraphics[scale=0.67]{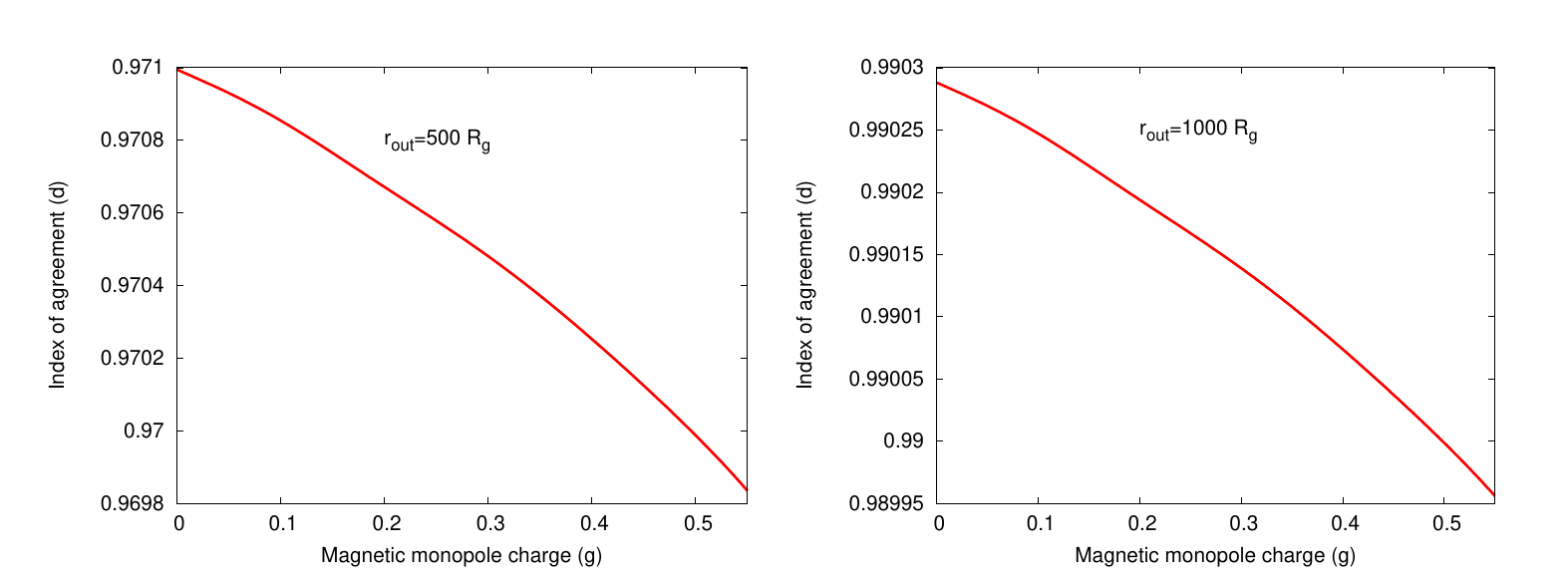}
\end{center}
\caption{Figure 5: The above figure shows the variation of the index agreement $d$ with the magnetic monopole charge parameter $g$. The left panel depicts the index of agreement computed with $r_{out}=500r_g$, while the right panel shows the index of agreement computed with $r_{out}=1000r_g$. Both peak at $g=0$ which confirms that the Kerr scenario is most favored.}
\label{indexd_Bardeen}
\end{figure}

The index of agreement $d$ \cite{willmott1984evaluation, doi:10.1080/02723646.1981.10642213,2005AdG.....5...89K} is proposed in order to overcome the insensitivity of Nash-Sutcliffe efficiency and its modified form towards the differences between the theoretical and the observed luminosities from the respective observed mean \cite{WRCR:WRCR8013} and its functional form is expressed as follows,

\begin{align}\label{eq_indexd_Bardeen}
&d\bigg(g, a_{min}, \cos i_{min},\dot{M}_{min} \bigg)=
\nonumber \\
&1-\frac{\sum_{j}\left\{\mathcal{O}_{j}-\Omega_{j}\bigg(g,\left \lbrace a_{min}, \cos i_{min},\dot{M}_{min} \right\rbrace \bigg)\right\}^{2}}{\sum _{j}\left\{ \big|\mathcal{O}_{j}-\mathcal{O}_{\rm av}\big|+\big|\Omega_{j}\bigg(g,\left \lbrace a_{min}, \cos i_{min},\dot{M}_{min}  \right\rbrace \bigg)-\mathcal{O}_{\rm av}\big|\right\}^{2}}
\end{align}
where, $\mathcal{O}_{\rm av}$ refers to the average value of the observed luminosities. The denominator, often known as the potential error, denotes the maximum deviation of each pair of observed and predicted luminosities from the average luminosity. 
\begin{figure}
\begin{center}
\hspace{-2.1cm}
\includegraphics[scale=0.67]{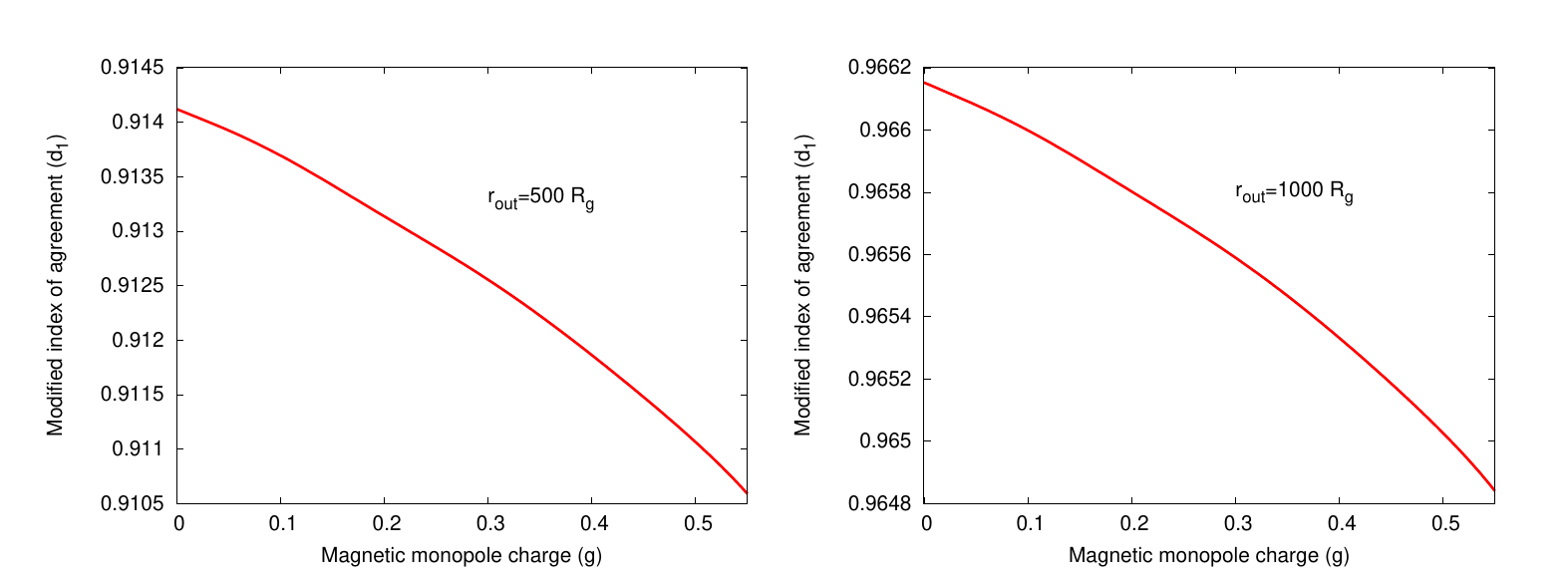}
\end{center}
\caption{Figure 6: The variation of the modified version of the index of agreement $d_1$ with magnetic monopole charge parameter $g$, computed with $r_{out}=500r_g$ (left panel) and $r_{out}=1000r_g$ (right panel) is shown in the figure above. Once again the modified index of agreement also maximises for $g=0$, regardless of the choice of $r_{out}$. }
\label{d1_Bardeen}
\end{figure}
Again due to the presence of squared terms in the numerator the index of agreement suffers from oversensitivity to higher values of optical luminosity and hence a modified version $d_1$ is proposed, where,
\begin{align}\label{eq_d1_Bardeen}
&d_{1}\bigg(g,\left \lbrace a_{min}, \cos i_{min},\dot{M}_{min} \right\rbrace \bigg)=\nonumber \\
&1-\frac{\sum_{j}\big|\mathcal{O}_{j}-\Omega_{j}\bigg(g,\left \lbrace a_{min}, \cos i_{min},\dot{M}_{min} \right\rbrace \bigg)\big|}{\sum_{j}\left\{\big|\mathcal{O}_{j}-\mathcal{O}_{\rm av}\big|+\big|\Omega_{j}\bigg(g,\left \lbrace a_{min}, \cos i_{min},\dot{M}_{min}\right\rbrace \bigg)-\mathcal{O}_{\rm av}\big| \right\}}
\end{align}

From \ref{eq_indexd_Bardeen} and \ref{eq_d1_Bardeen} it is clear that the choice of $g$ for which $d$ and $d_1$ maximizes is the most favored monopole charge derived from the observed sample. It is important to note that both the estimators $d$ and $d_1$ maximizes for $g=0$ which again is consistent with our previous results. Also, irrespective of the choice of $r_{out}$, the most favored magnetic monopole charge parameter turns out to be $g=0$ which is validated from all the five error estimators discussed here.

\end{itemize}

The entire analysis of the error estimators reveal that the $\chi^2$ minimizes for $g=0$ while the other error estimators maximize for the same magnetic monopole charge parameter value. It is interesting to note that estimating the error estimators for different choice of $r_{out}$ keep the result unaltered. Since the quasars are rotating in nature, it is important to report the spins of the same corresponding to the most favored choice of the monopole parameter $g$ which in turn corresponds to the Kerr scenario in \gr.

\subsection{Spins of the quasars}\label{S4_Bardeen-2}
In this section we try to estimate the observationally favored values of spin from the quasar optical data. We have discussed earlier in \ref{S4_Bardeen} the procedure to extract the most favored value of spin for a given $g$. Since $g=0$ (which corresponds to the Kerr scenario) is observationally most favored we report in \ref{Table1_Bardeen} the values of spin associated with $g=0$. For some of these quasars the spins are estimated by other independent methods which will be compared with our present findings in this section.
In \ref{S4_Bardeen-1} it has already been discussed that the theoretical estimates of optical luminosity, $L_{opt}$ depends both on the marginally stable orbit radius $r_{ms}$ as well as the outer radius of the disk, $r_{out}$. However, the flux emitted from the accretion disk peaks close to $r_{ms}$ and hence the emission from the inner radius has much greater impact on the luminosity than the outer parts of the disk. Therefore, the choice of outer radius should not significantly affect our findings. This is further confirmed by our results in the last section where the observationally favored value of $g$ does not change by varying the outer radius from $r_{out}=500r_{g}$ to $r_{out}=1000r_g$ although the value of error estimators depend on $r_{out}$.

\begin{table}

\caption{Table 1: Spin parameters of quasars corresponding to $g=0$}
\label{Table1_Bardeen}
\vspace{-0.36cm}
\begin{center}
\centering
\begin{tabular}{|c|c|c|c|c|}

\hline
$\rm Object$ & $\rm log~ m$ &  $\rm log ~L_{obs}$ & $\rm log ~L_{bol}$ & $a_{g=0}$\\
\hline
$\rm 0003+199$ & $\rm 6.88$ &  $\rm 43.91$ & $\rm 45.13 \pm 0.35$ & $\rm 0.99 $\\ \hline
$\rm 0050+124$ & $\rm 6.99$  & $\rm 44.41$ & $\rm 45.12 \pm 0.04$ & $\rm 0.99 $\\ \hline
$\rm 0923+129$ & $\rm 6.82$  & $\rm 43.58$ & $\rm 44.53 \pm 0.15$ & $\rm 0.99 $\\ \hline
$\rm 1011-040$ & $\rm 6.89$ & $\rm 44.08$ & $\rm 45.02 \pm 0.23$ & $\rm 0.99 $\\ \hline
$\rm 1022+519$ & $\rm 6.63$ & $\rm 43.56$ & $\rm 45.10 \pm 0.39$ & $\rm 0.99$\\ \hline
$\rm 1119+120$ & $\rm 7.04$ & $\rm 44.01$ & $\rm 45.18 \pm 0.34$ & $\rm 0.99 $\\ \hline
$\rm 1244+026$ & $\rm 6.15$ & $\rm 43.70$ & $\rm 44.74 \pm 0.22$ & $\rm 0.99 $\\ \hline
$\rm 1404+226$ & $\rm 6.52$ &  $\rm 44.16$ & $\rm 45.21 \pm 0.26$ & $\rm 0.99 $\\ \hline
$\rm 1425+267$ & $\rm 9.53$ &  $\rm 45.55$ & $\rm 46.35 \pm 0.20$ & $\rm 0.7 $\\ \hline
$\rm 1440+356$ & $\rm 7.09$ & $\rm 44.37$ & $\rm 45.62 \pm 0.29$ & $\rm 0.99 $\\ \hline
$\rm 1535+547$ & $\rm 6.78$ & $\rm 43.90$ & $\rm 44.34 \pm 0.02$ & $\rm 0.99 $\\ \hline
$\rm 1545+210$ & $\rm 9.10$ &  $\rm 45.29$ & $\rm 46.14 \pm 0.13$ & $\rm 0.1 $\\ \hline
$\rm 1552+085$ & $\rm 7.17$  & $\rm 44.50$ & $\rm 45.04 \pm 0.01$ & $\rm 0.99 $\\ \hline
$\rm 1613+658$ & $\rm 8.89$  & $\rm 44.75$ & $\rm 45.89 \pm 0.11$ & $\rm 0.3 $\\ \hline
$\rm 1704+608$ & $\rm 9.29$ &  $\rm 45.65$ & $\rm 46.67 \pm 0.21$ & $\rm -0.4 $\\ \hline
$\rm 2308+098$ & $\rm 9.43$ & $\rm 45.62$ & $\rm 46.61 \pm 0.22$ & $\rm 0.95 $\\ \hline
\end{tabular}
\end{center}
\end{table}

\par
In \ref{Table1_Bardeen}, the spins of quasars which remain independent of the choice of $r_{out}$ is reported. It is important to note that the choice of $r_{out}$ do affect the best choice of spins corresponding to a given $g$ for some quasars in our model. This is because we allowed variation of accretion rates and inclination angle for each of the quasars. Further, apart from the metric parameters the theoretical optical luminosity which in turn is related to the temperature profile $T(r)$ depends on the ratio $\dot{M}/\mathcal{M}^2$. Hence, although the optical luminosity generally has negligible contribution from $r_{out}$, the quasars for which this ratio is high the outer disk also contributes to some extent to the optical luminosity. In \ref{Table1_Bardeen} we report the spins of only those quasars where the most favored choice of spin $a_{min}$ remains unaltered with $r_{out}$.
It turns out that for most of the quasars whose spin remain unchanged with $r_{out}$ are maximally spinning with $a \sim 0.99$ for $g=0$ (see \ref{Table1_Bardeen}). 

We now compare the spins obtained from the present analysis with earlier findings. According to \cite{Ross:2005dm}, \cite{Crummy:2005nj}, PG 0003+199, PG 0050+124, PG 1244+026, PG 1404+226, PG 1440+356 are maximally spinning which is based on the general relativistic disk reflection model. Keek et al. \cite{Keek:2015apa} also constrained the spin of PG 0003+199 to $a \sim 0.89\pm 0.05$ while according to the findings of Walton et al. \cite{Walton:2012aw} the spin of PG 0003+199 turns out to be $a\sim 0.83^{+0.09}_{-0.13}$. These results are nearly consistent with our estimated spins (\ref{Table1_Bardeen}).
From the iron-line method Bottacini et al. \cite{Bottacini:2014lva} reported that PG 1613+658 (Mrk 876) harbors a rotating black hole which is in agreement with our findings. The spins of PG 0003+199, PG 0050+124, PG 0923+129, PG 2308+098, PG 1022+519, PG 1425+267, PG 1545+210, PG 1613+658 and PG 1704+608 have been independently estimated based on polarimetric observations of AGNs \cite{Afanasiev:2018dyv}. While the spins of PG 0003+199, PG 0050+124, PG 0923+129 and PG 2308+098 are in agreement with our results, the spins of PG 1022+519, PG 1425+267, PG 1545+210, PG 1613+658 and PG 1704+608 exhibit some variations. Moreover, \cite{2017Ap&SS.362..231P} constrained the spin of PG 1704+608 (3C 351) to $a<0.998$ in tandem with our results.

\section{Concluding Remarks}\label{S5_Bardeen}
In this paper we derive the continuum spectrum from the accretion disk around black holes in Bardeen spacetime. The significance of the Bardeen spacetime lies in the fact that it can avoid the curvature singularity in black holes by introducing a magnetic monopole charge. Thus, the Bardeen spacetime is characterized by the rotation parameter $a$ and the magnetic monopole charge parameter $g$. After computing the theoretical spectrum in the Bardeen spacetime we compare it with the optical observations of eighty Palomar Green quasars. Our analysis reveals that the Kerr scenario in \gr\ is observationally more favored than black holes with a monopole charge. However, it may be noted that the Kerr scenario also arises in other theories of gravity apart from \gr\ \cite{Psaltis:2007cw}. 
Our results are obtained by computing error estimators like the $\chi^2$, the Nash-Sutcliffe efficiency, the index of agreement and their modified forms. We also report the spins corresponding to some of the quasars for $g=0$. We note that magnetic monopole charge with $ g\geq 0.03$ is outside $99\%$ confidence interval.

Before concluding we would like to mention several limitations of the present analysis. First, the spectral energy distribution (SED) of the quasars comprises of the accretion disk, the corona, the jet and the dusty torus which are not always easy to observe and model \cite{Brenneman:2013oba}. Understanding the role of each of these components e.g. the accretion disk, the corona, the jet and the dusty torus
on the SED is extremely difficult which limits accurate determinination of the black hole parameters, e.g. distance, inclination, mass, spin or other associated metric components. As a result the spin of the same
quasar estimated by different methods assuming \gr\ often leads to inconsistent results \cite{Brenneman:2013oba,Reynolds:2013qqa,Steiner:2012vq,Gou:2009ks,Reynolds:2019uxi}.

Second, the continuum spectrum depends not only on the background spacetime but also on the properties of the accretion flow.
In the present work the spectrum is computed using the Novikov-Thorne model where the effects of outflows or the radial velocity of the accretion flow are not taken into account. A more comprehensive modelling of the disk is expected to constrain the background spacetime better. Presently, these issues are addressed by several phenomenological models which is beyond the scope of this work. Further, it may be worthwhile to carry out the present analysis with a different observational sample of black holes.
Apart from the continuum spectrum, observations like the black hole shadow \cite{Akiyama:2019brx,Akiyama:2019fyp,Akiyama:2019eap}, quasi-periodic oscillations \cite{Maselli:2014fca,Pappas:2012nt} and iron $K-\alpha$ line \cite{Bambi:2016sac,Ni:2016uik} in the power spectrum and the reflection spectrum of black holes respectively are some of the independent observations which can be used to probe the background metric. Such a study will be conducted in future and reported elsewhere.

\section*{Acknowledgements}
The research of SSG is partially supported by the Science and Engineering
Research Board-Extra Mural Research Grant No. (EMR/2017/001372), Government of India.
The authors thank Dr. Sumanta Chakraborty for useful comments and discussions throughout the course of this work.

\bibliography{accretion,accretion2,torsion,Gravity_1_full,Gravity_2_full,Gravity_3_partial,Brane,KN-ED,KN-ED2,EMDA-Jet,bardeen,Black_Hole_Shadow}

\bibliographystyle{./utphys1}

\end{document}